\documentclass[manuscript]{emulateapj}

\shorttitle{Magnetic reconnection between small-scale loops}
\shortauthors{Yang et al.}

\journalinfo{Accepted for publication in ApJL}

\submitted{Accepted for publication in ApJL}

\begin{document}

\title{Magnetic reconnection between small-scale loops observed with the New Vacuum Solar Telescope}

\author{Shuhong Yang\altaffilmark{1}, Jun Zhang\altaffilmark{1}, and
Yongyuan Xiang\altaffilmark{2}}

\altaffiltext{1}{Key Laboratory of Solar Activity, National
Astronomical Observatories, Chinese Academy of Sciences, Beijing
100012, China; shuhongyang@nao.cas.cn}

\altaffiltext{2}{Fuxian Solar Observatory, Yunnan Observatories,
Chinese Academy of Sciences, Kunming 650011, China}

\begin{abstract}

Using the high tempo-spatial resolution H$\alpha$ images observed
with the New Vacuum Solar Telescope, we report the solid
observational evidence of magnetic reconnection between two sets of
small-scale anti-parallel loops with an X-shaped topology. The
reconnection process contains two steps: a slow step with the
duration of more than several tens of minutes, and a rapid step
lasting for only about three minutes. During the slow reconnection,
two sets of anti-parallel loops reconnect gradually, and new loops
are formed and stacked together. During the rapid reconnection, the
anti-parallel loops approach each other quickly, and then the rapid
reconnection takes place, resulting in the disappearance of former
loops. In the meantime, new loops are formed and separate. The
region between the approaching loops is brightened, and the
thickness and length of this region are determined to be about 420
km and 1.4 Mm, respectively. During the rapid reconnection process,
obvious brightenings at the reconnection site and apparent material
ejections outward along reconnected loops are observed. These
observed signatures are consistent with predictions by reconnection
models. We suggest that the successive slow reconnection changes the
conditions around the reconnection site and triggers instabilities,
thus leading to the rapid approach of the anti-parallel loops and
resulting in the rapid reconnection.

\end{abstract}


\section{INTRODUCTION}

Magnetic reconnection is a rearrangement of magnetic field topology,
and it is a fundamental physical process in conductive plasma.
Magnetic flux is frozen in the plasma except in the small diffusion
region where magnetic reconnection takes place (see Zweibel \&
Yamada 2009; Yamada et al. 2010). When magnetic field lines
reconnect, magnetic energy is converted to the thermal energy and
kinetic energy of plasmas. According to most of the theories, in the
magnetic reconnection, there should exist a small dissipation
region, topological changes, strong outflows, and other signatures
of magnetic energy release, such as sudden brightenings (e.g.,
Parker 1957; Sweet 1958; Furth et al. 1963; Petschek 1964). Magnetic
reconnection is often considered to be the mechanism driving energy
release in solar flares, stellar flares, and many types of jets and
outbursts (Rosner et al. 1985; Haisch et al. 1991; Yuan et al.
2009).

The evidences of magnetic reconnection have been observed in
different types of solar events, such as flares, coronal mass
ejections, and solar wind (Masuda et al. 1994; Gosling et al. 2007;
Li \& Zhang 2009). Especially, as one of the most energetic
phenomena on the Sun, solar flares are widely deemed to be caused by
the successive reconnection of magnetic field lines in the corona.
Many signatures of magnetic reconnection during flares have been
observed, such as cusp-shaped structures above flare loops (Tsuneta
et al. 1992), the transformation of ``open" loops to closed
post-flare ones (Zhang et al. 2013), the reconnection inflows and
outflows (Yokoyama et al. 2001; Innes et al. 2003; Asai et al. 2004;
Lin et al. 2005; Takasao et al. 2012; Su et al. 2013). In situ
measurements also revealed the occurrence of magnetic reconnection
in the magnetosphere (Mozer et al. 2002; Phan et al. 2007; Xiao et
al. 2006, 2007; Dunlop et al. 2011). When the magnetic fields
carried by solar wind travel outward from the Sun and interact with
the planetary magnetic fields, current sheets are created and
magnetic reconnection occurs. In laboratories, experiments dedicated
to magnetic reconnection have been extensively carried out under
controlled conditions (Bratenahl \& Yeates 1970; Yamada et al.
1997). With intense lasers in the laboratory, Zhong et al. (2010)
reconstructed a magnetic reconnection topology which is similar to
that in solar flares. In their experiment, loop-top-like X-ray
source emission and outflows were reproduced successfully and the
diffusion regions were also identified.

As the primary observing facility of the \emph{Fuxian Solar
Observatory} in China, the New Vacuum Solar Telescope (NVST; Liu et
al. 2014), a vacuum telescope with a clear aperture of 985 mm, is
designed to observe the Sun at high temporal and spatial
resolutions. The H$\alpha$ line which is formed in the chromosphere
is quite useful to investigate the fine structures of dynamic
events. In this Letter, using the NVST H$\alpha$ images, we report a
well observed process of magnetic reconnection with two steps
between small-scale loops in the chromosphere, which provides
observational evidence of magnetic reconnection as predicted in
theories.

\section{OBSERVATIONS AND DATA ANALYSIS}

\begin{figure*}
\centering
\includegraphics
[bb=100 176 490 653,clip,angle=0,width=0.8\textwidth] {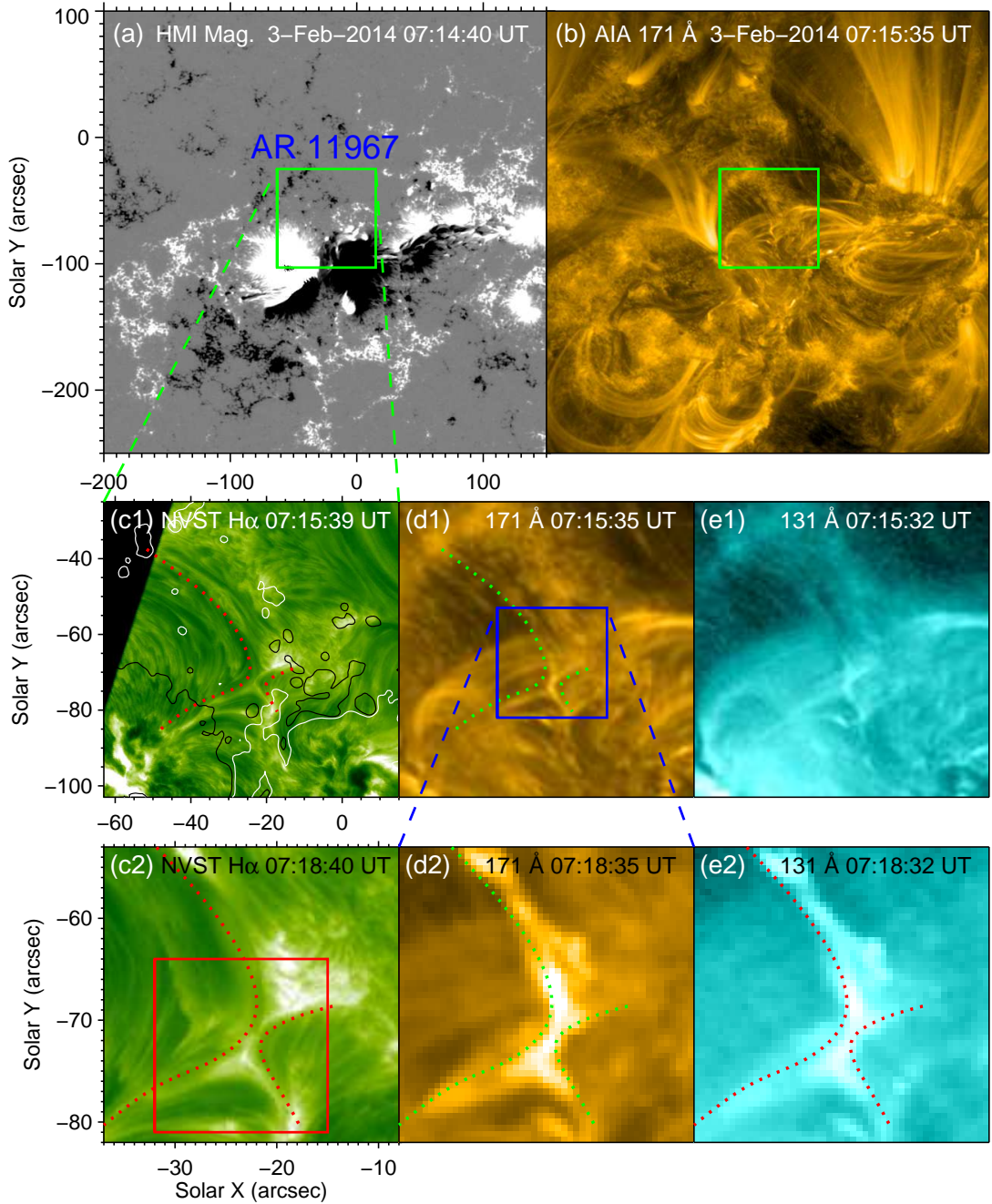}
\caption{Panels (a)-(b): HMI line-of-sight magnetogram and AIA 171
{\AA} image displaying the overview of AR 11967. Panels (c1)-(e1):
NVST H$\alpha$, AIA 171 {\AA}, and 131 {\AA} images showing the
expanded view of the area outlined by the square in panel (a).
Panels (c2)-(e2): similar to panels (c1)-(e1), but for the area
outlined by the square in panel (d1) 3 min later. The red square in
panel (c2) outlines the FOV of Figure 2. The black and white curves
in panel (c1) are the contours of the positive and negative magnetic
fields at levels of 220 G and -220 G, respectively. The red and
green dotted curves outline two sets of loops involved in
reconnection identified in the H$\alpha$ image. \label{fig}}
\end{figure*}

\begin{figure*}
\centering
\includegraphics
[bb=62 235 503 590,clip,angle=0,width=0.9\textwidth] {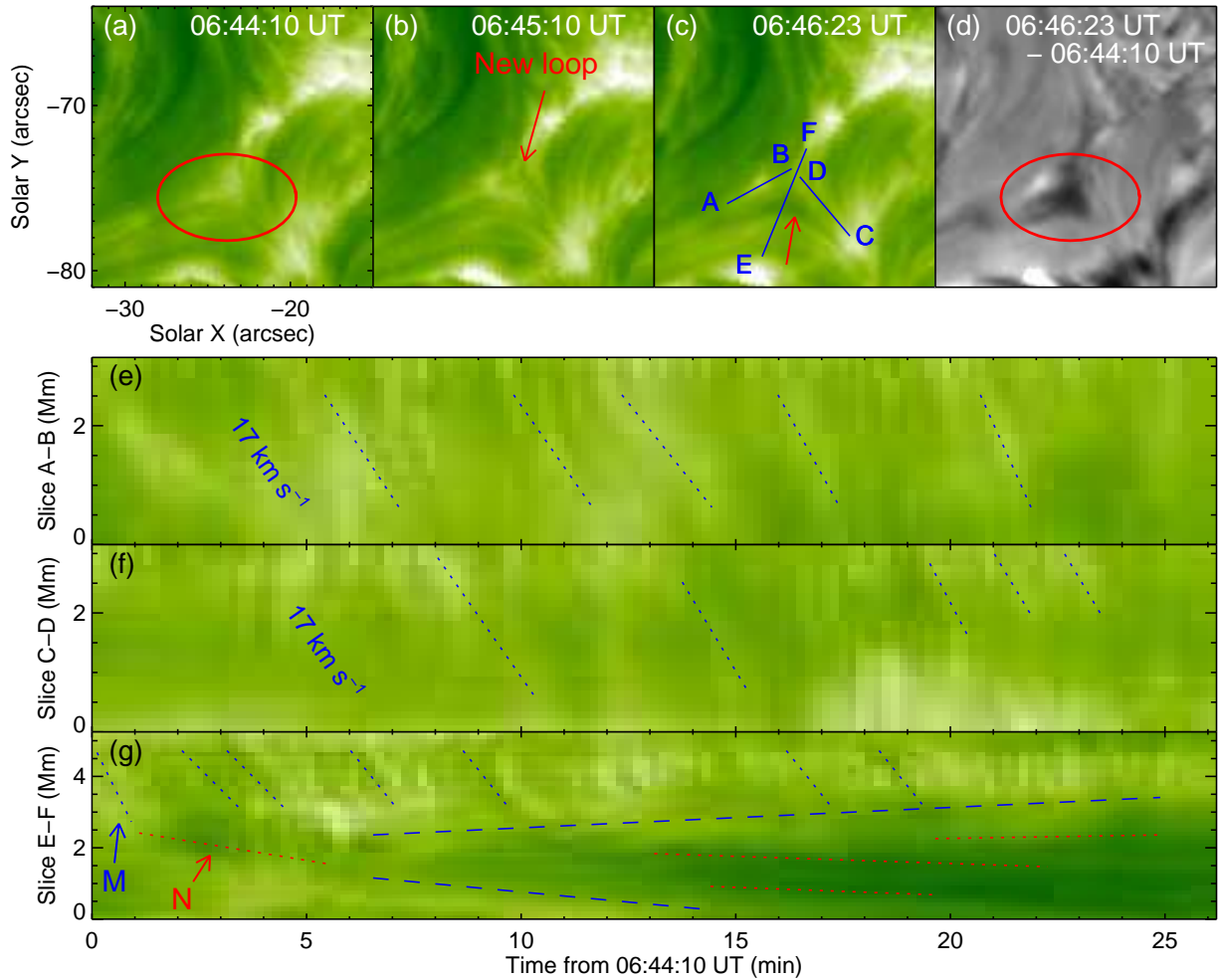}
\caption{ Panels (a)-(c): Sequence of H$\alpha$ images showing the
formation of a small loop during the slow reconnection process (also
see Movie 1). The arrows denote the newly formed loop. Panel (d):
difference image between 06:44:10 UT and 06:46:23 UT. The ellipses
in panels (a) and (d) outline the area with newly formed H$\alpha$
loop. Panels (e)-(g): space-time plots along slices ``A--B",
``C--D", and ``E--F", respectively, marked in panel (c). The dotted
lines in panels (e) and (f) mark the bright moving features, and
those in panel (g) follow the dark features. The dashed lines mark
the boundaries of the stack of newly formed loops. Arrows ``M" and
``N" denote the formation and retraction of the H$\alpha$ loop shown
in panels (a)-(c). \label{fig}}
\end{figure*}

\begin{figure*}
\centering
\includegraphics
[bb=48 175 538 656,clip,angle=0,width=0.9\textwidth] {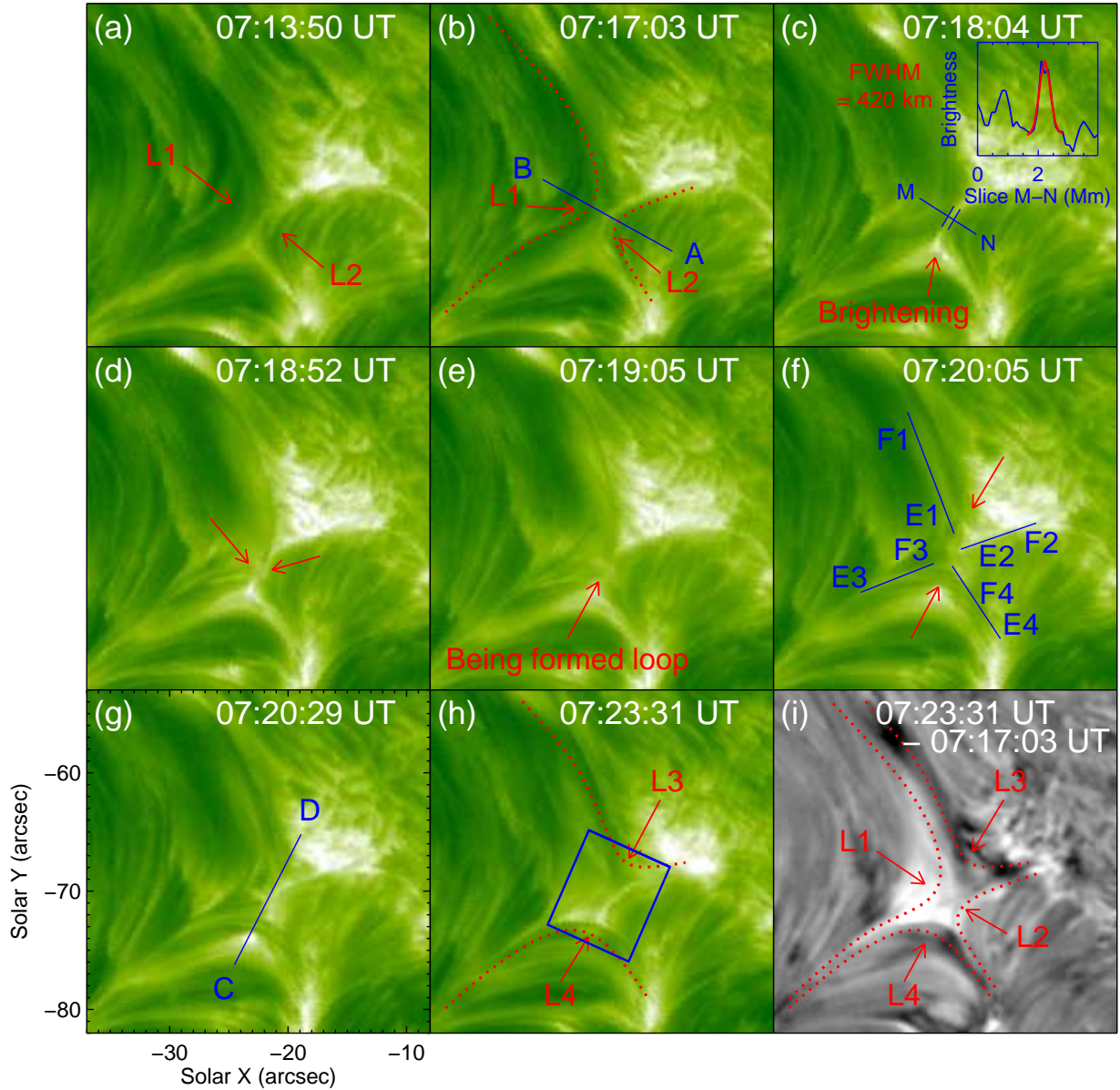}
\caption{Panels (a)-(h): sequence of H$\alpha$ images showing the
process of the rapid reconnection (also see Movie 2). Panel (i):
difference image between 07:17:03 UT and 07:23:31 UT. Arrows ``L1"
and ``L2" denote the to-be-reconnected loops before reconnection,
and arrows ``L3" and ``L4" denote the newly formed loops after
reconnection. The arrow in panel (c) indicates a brightening region,
and the two parallel lines mark the reconnection region of
interacting loops. The two arrows in panel (d) denote the break
points of loops ``L1" and ``L2", and the arrows in panels (e) and
(f) indicate the being formed loops. The blue square outlines the
area where the light curves shown in Figure 4(a) are derived, and
the blue lines mark the positions where the space-time plots
displayed in Figures 4(b)-(g) are obtained. \label{fig}}
\end{figure*}

\begin{figure}
\centering
\includegraphics
[bb=168 110 426 710,clip,angle=0,width=0.4\textwidth] {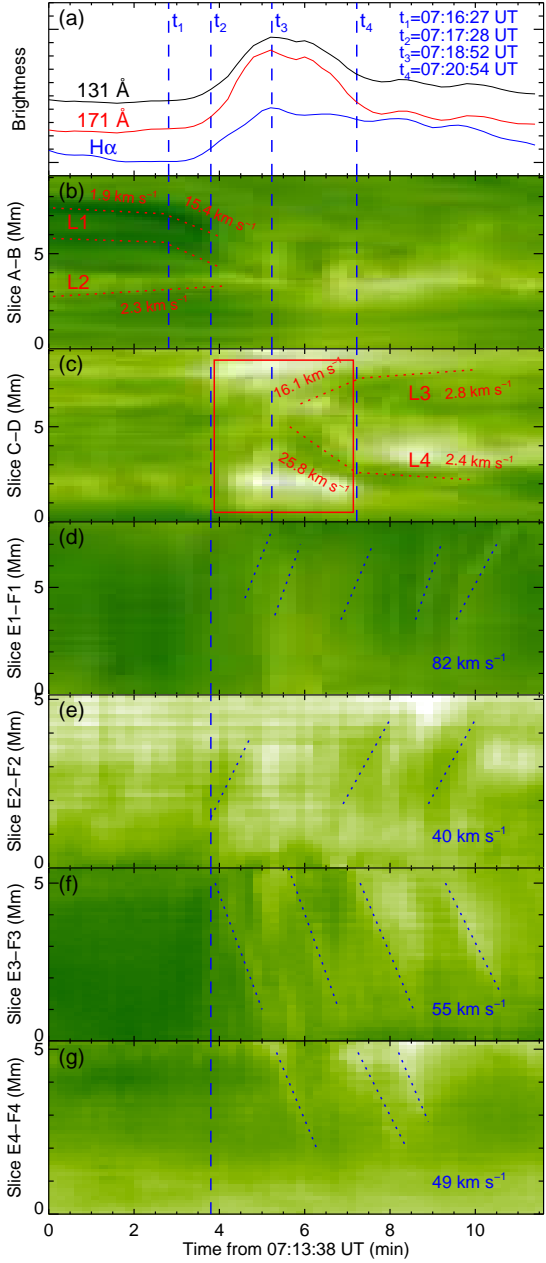}
\caption{Panel (a): light curves in H$\alpha$, 171 {\AA}, and 131
{\AA} lines derived from the area outlined by the square in Figure
3(h). Panels (b)-(g): space-time plots along slices ``A--B",
``C--D", ``E1--F1", ``E2--F2", ``E3--F3", and ``E4--F4",
respectively, marked in Figure 3. The dotted lines ``L1" and ``L2"
display the evolution of the loops before reconnection, and ``L3"
and ``L4" denote that of the new loops. The blue dotted lines in
panels (d), (f), and (g) follow the bright features, and those in
panel (e) follow the dark features. The vertical dashed lines label
the different stages of the rapid reconnection. The rectangle in
panel (c) outlines the brightenings during the reconnection process.
\label{fig}}
\end{figure}

\begin{figure*}
\centering
\includegraphics
[bb=60 385 533 466,clip,angle=0,width=0.95\textwidth] {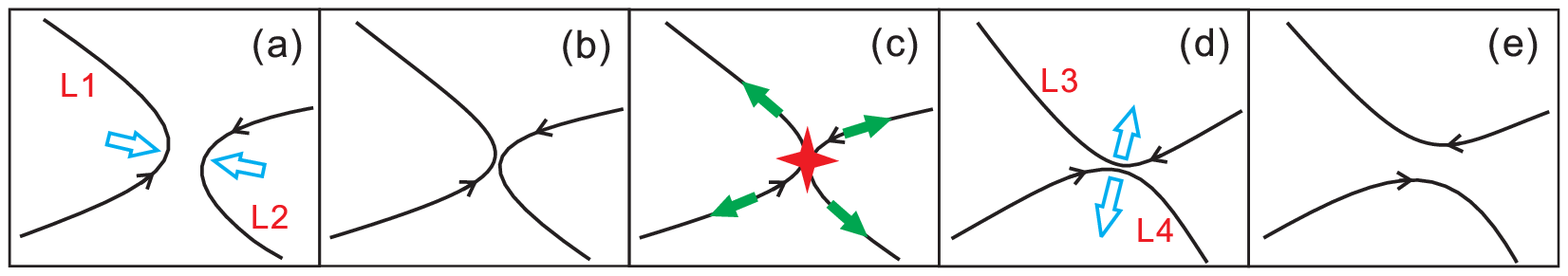}
\caption{Schematic drawings illustrating the magnetic reconnection
process observed in this study. The blue arrows in panel (a)
indicate the convergence of loops ``L1" and ``L2", and the blue
arrows in panel (d) indicate the separation of the newly formed
loops ``L3" and ``L4". In panel (c), the red symbol marks the
reconnection site, and the green arrows represent the material
ejections due to the reconnection. \label{fig}}
\end{figure*}

On 2014 February 3, the NVST was pointed to AR 11967 with a
field-of-view (FOV) of 151$''$ $\times$ 151$''$. The NVST data used
in this study were obtained in H$\alpha$ 6562.8 {\AA} line from
05:49:52 UT to 09:10:01 UT. The H$\alpha$ images have a cadence of
12 s and a pixel size of 0$\arcsec$.163. The data are calibrated
from Level 0 to Level 1 with dark current subtracted and flat field
corrected, and then the calibrated images are reconstructed to Level
1+ by speckle masking (Weigelt 1977; Lohmann et al. 1983). In
addition, the Atmospheric Imaging Assembly (AIA; Lemen et al. 2012)
multi-wavelength images and the Helioseismic and Magnetic Imager
(HMI; Scherrer et al. 2012; Schou et al. 2012) line-of-sight
magnetograms from the \emph{Solar Dynamics Observatory} (\emph{SDO};
Pesnell et al. 2012) are also adopted. We choose the AIA 171 {\AA}
and 131 {\AA} images to study the process of magnetic reconnection
at different temperatures. The AIA images were obtained from 05:30
UT to 09:30 UT on February 3 with a pixel size of 0$\arcsec$.6 and a
cadence of 12 s. We use the HMI line-of-sight magnetograms observed
from 00:00 UT on February 1 to 00:00 UT on February 5. They have a
spatial sampling of 0$\arcsec$.5 pixel$^{-1}$ and a cadence of 45 s.
The AIA and HMI data are calibrated to Level 1.5 by using the
standard procedure \emph{aia\_prep.pro}, and rotated differentially
to a reference time (07:15:00 UT on February 3). Then, we co-align
the \emph{SDO} and NVST images using the cross-correlation method
with specific features.

\section{RESULTS}

Magnetic reconnection took place at the edge of AR 11967 (see Figure
1(a)). At 07:15 UT, there was a small X-shaped structure at the
reconnection site in the AIA 171 {\AA} image (outlined by the green
square in panel (b)), which can be identified more clearly in the
expanded view (panel (d1)). In the AIA 131 {\AA} band this structure
was also conspicuous (panel (e1)). The X-shaped structure observed
in the EUV images was located between two set of loops (outlined by
the dotted curves) identified in the H$\alpha$ image (panel (c1)).
It should be mentioned that, as noted by Yang et al. (2014),
magnetic loops can be indicated by the dark fibrils in H$\alpha$
images. Here, the reconnection occurred between the loops outlined
by the dotted curves. The left loops connected the positive sunspot
and the nearby negative fields, and the right loops linked the
negative sunspot with the nearby small-scale fields of positive
polarity. Three minutes later, the brightness of the X-shaped
structure increased significantly in 171 {\AA} and 131 {\AA} images
(see panels (d2) and (e2)), and there were also some brightenings
and changes of H$\alpha$ fibrils (panel (c2)). The reconnection
process can be divided into two steps: a slow step followed by a
rapid one.

\subsection{Step one: slow reconnection}

In the area outlined by the red square in Figure 1(c2), slow
reconnection was observed for several tens of minutes, and part of
the reconnection is shown in Movie 1. Figures 2(a)-(c) display the
formation of a small loop during the slow reconnection process. In
our observations, the high-density and low-temperature plasmas may
be not sufficiently heated, i.e., some plasmas are heated to
high-temperature while others are still at low-temperature.
Therefore, newly formed loops can be outlined by either bright or
dark features in the outflow regions. In the ellipse region (panel
(a)), a being formed loop could be identified at 06:45:10 UT
(denoted by the red arrow in panel (b)). At 06:46:23 UT, the newly
formed loop was much clearer and reached to a lower site (denoted by
the red arrow in panel (c)). In the difference image (panel (d))
between 06:44:10 UT and 06:46:23 UT, there exists a dark structure
as outlined by the ellipse, indicating the formation of dark loop.

Along slice ``A-B", ``C-D", and ``E-F" marked in panel (c), we make
three space-time plots and display them in panels (e)-(g),
respectively. We can see that, there are many bright features, which
seem to be the heated blob-like plasmas, moving from ``B" to ``A",
and from ``D" to ``C" (as indicated by dotted lines in panels (e)
and (f)). The average velocity of the apparent motion of bright
features both along ``B-A" and ``D-C" is about 17 km s$^{-1}$. In
panel (g), the line denoted by arrow ``M" indicates the quick
downward motion of the newly formed loop shown in panel (b). After
the quick motion, the new loop continued to move downward slowly, as
denoted by arrow ``N". As the reconnection went on, more and more
new loops were formed and stacked together (see the other blue and
red dotted lines in panel (g)). The boundaries of the loop stack are
marked by the dashed lines.

\subsection{Step two: rapid reconnection}

Figure 3 shows the process of the rapid reconnection (also see Movie
2). The loops involved into the reconnection are labeled with ``L1"
and ``L2", as denoted by the arrows in panels (a) and (b). Before
the occurrence of the rapid reconnection, the two sets of loops were
approaching each other. At 07:17:03 UT, the distance between loops
``L1" and ``L2" was much shorter than that at 07:13:50 UT. Then
loops ``L1" and ``L2" continued to move to each other and eventually
interacted. At 07:18:04 UT, an obvious brightening (denoted by the
arrow in panel (c)) can be observed. The two short parallel lines
mark the reconnection region of the interacting loops. The inserted
blue curve shows the brightness along slice ``M-N", and the overlaid
red curve is the Gaussian fit of the brightness profile. The width
of the bright interaction region is given by the width of the
Gaussian, which is found to be 420 km. The length (presented by the
length of the parallel lines) of the interaction region is about 1.4
Mm. At 07:18:52 UT, both loops ``L1" and ``L2" broke apparently (as
denoted by the two arrows in panel (d)). Only 13 s later, a new loop
had been formed, as denoted by the arrow in panel (e). At 07:20:05
UT, the new loop was more conspicuous (indicated by the lower arrow
in panel (f)) and another new loop (denoted by the higher arrow) can
also be observed. Then the two new loops retracted and moved away
from each other (panels (g)-(h)). The difference image between
07:17:03 UT and 07:23:31 UT is displayed in panel (i). In the
difference image, the locations of the former loops ``L1" and ``L2"
are white structures, which is caused by the disappearance of loops
``L1" and ``L2". While the black structures are coincided with the
locations of loops ``L3" and ``L4", indicating the formation of new
H$\alpha$ loops.

During the rapid reconnection process, the most remarkable changes
are the disappearance of loops ``L1" and ``L2" and the formation of
loops ``L3" and ``L4". To clearly display these changes with time,
we make two space-time plots along slices ``A-B" and ``C-D" which
are marked in Figure 3, and display them in Figures 4(b) and (c).
When the reconnection began, we can find apparent ejections of
bright features outward along ``E1-F1", ``F3-E3" and ``F4-E4", and
dark features along ``E2-F2" (marked in Figure 3) from the
reconnection site. It should be noted that, as displayed in Figure
3(f), the background of slice ``E2-F2" is so bright that the
expected bright moving features could not be distinguished, while
some dark features can be identified. Then we derive four space-time
plots along the four slices and exhibit them in Figures (d)-(g),
respectively. We also measure the brightness variations of
H$\alpha$, 171 {\AA}, and 131 {\AA} images in the region outlined by
the rectangle in Figure 3(h), and the corresponding light-curves are
shown in Figure 4(a). Before 07:16:27 UT ($t_{1}$, indicated by the
leftmost vertical line in Figure 4), loop ``L1" approached loop ``2"
slowly with an average velocity of 1.9 km s$^{-1}$ (see panel (b)).
While after $t_{1}$, loop ``L1" moved toward loop ``L2" quickly with
a velocity of 15.4 km s$^{-1}$, and contacted each other at 07:17:28
UT ($t_{2}$, the second vertical line). Before $t_{2}$, there was no
obvious variation for the brightness in each wavelength (panel (a)),
and no significant moving feature along different directions (panels
(d)-(g)). After $t_{2}$, when loop ``L1" began to interact with loop
``L2" (panel (b)), the brightness in each wavelength increased
rapidly (see panel (a)). At and around the reconnection region, the
brightenings in H$\alpha$, 171 {\AA}, and 131 {\AA} lines can be
found in panels (a) and (c) and Movie 2. All the three light curves
reach the maximum at the same time, i.e., 07:18:52 UT ($t_{3}$ in
panel (a)). Also at that time, loops ``L1" and ``L2" disappeared
(see panel (b)) and loops ``L3" and ``L4" began to be formed (panel
(c)). From $t_{3}$, the light curves exhibit a decrease trend (panel
(a)), and the newly formed loops separated quickly with a separation
velocity of 42 km s$^{-1}$ (panel (c)). When the reconnection ended
at around 07:20:54 UT ($t_{4}$), the brightenings in
multi-wavelengths almost disappeared (see panels (a) and (c) and
Movie 2). The separation velocity also slowed down to about 5.2 km
s$^{-1}$ (as indicated by the dotted curves in panel (c)). In panels
(d)-(g), moving features only can be identified after $t_{2}$, i.e.,
the start time of the rapid reconnection. The apparent motions of
small features along all of the four directions can be observed, and
they have a comparable velocity of about 50 km s$^{-1}$.

\section{CONCLUSIONS AND DISCUSSION}

Using the NVST H$\alpha$ images with high tempo-spatial resolutions,
we have observed signatures of magnetic reconnection between two
sets of small-scale loops. The reconnection process can be divided
into two steps: a slow reconnection with the duration of several
tens of minutes and a rapid reconnection lasting for only about
three minutes. During the slow reconnection process, two sets of
anti-parallel loops gradually reconnected, and new loops were formed
and stacked together. During the rapid reconnection, the
anti-parallel loops moved toward each other quickly, and then the
rapid reconnection took place, resulting in the disappearance of
former loops. In the meantime, new loops were formed and separated.
During the rapid reconnection process, we have observed obvious
brightenings at the reconnection site and apparent ejections of
bright or dark features outward along the newly formed loops with an
average velocity of about 50 km s$^{-1}$.

According to the observational results, we sketch a series of
cartoons (see Figure 5) to illustrate the reconnection process. The
loops ``L1" and ``L2" which will be reconnected move toward each
other to a very close distance, as shown in panels (a) and (b). When
the loops are close enough (about 420 km determined in this study),
magnetic reconnection between them takes place. At the reconnection
site, brightenings are observed, and along different directions,
apparent ejections of small features outward from the reconnection
site can be observed (see panel (c)). Then two new loops ``L3" and
``L4" are formed while the former loops ``L1" and ``L2" have
disappeared, and the newly formed loops begin to separate (panels
(d) and (e)). The common features of most reconnection theories
include the changes of magnetic topologies and the release of
magnetic energy (Parker 1957; Sweet 1958; Petschek 1964; see the
review by Yamada et al. 2010). The topology changes are mainly the
break of inflowing anti-parallel loops and the formation of new
loops. When the loops reconnect in the diffusion region, magnetic
energy is released, thus heating the plasmas. The plasma pressure is
raised, and a great deal of energy is converted to the kinetic
energy. In addition, the reconnected field lines near the X-point
are sharply bent and the magnetic tension force also impacts on the
plasmas to increase the kinetic energy. Therefore, the plasmas are
brightened and expelled. All the above features which should appear
in the magnetic reconnection have been observed in the present
study. Our results are highly consistent with the common models of
magnetic reconnection.

Until now, many evidences of magnetic reconnection on the Sun have
been reported by many authors (Tsuneta et al. 1992; Yokoyama \&
Shibata 1995; Yang et al. 2011; Takasao et al. 2012; Cirtain et al.
2013). However, most of these observational signatures of magnetic
reconnection have been found in solar eruptive events. According to
the popular flare model, a rising flux rope in the corona stretches
the overlying magnetic fields lines, and a current is created
between the anti-parallel field lines and magnetic reconnection
occurs (Shibata et al. 1995; Tsuneta 1996; Lin \& Forbes 2000). In
the present study, the magnetic reconnection is observed in a
relative stable X-shaped structure in the chromosphere compared with
that observed during flares in the corona. The previous observations
have revealed that the current sheets have a thickness $>$ 10$^{4}$
km and a length of more than several hundreds of Mm (Ciaravella \&
Raymond 2008; Lin et al. 2009). In the present study, the thickness
and length of the brightening region between the approaching loops
are only about 420 km and 1.4 Mm, respectively. It is likely that a
current sheet is embedded inside this structure of enhanced
emission. If so, the current sheet determined in this study is much
smaller than those previously reported.

In our observations, the reconnection process includes two steps,
i.e., a slow reconnection followed by a rapid reconnection. The slow
reconnection lasted for several tens of minutes, while the rapid
step only took about three minutes. We suggest that the continual
slow reconnection changed the conditions around the reconnection
region which triggered instabilities and led to rapid approaching of
the anti-parallel loops, thus resulting in the rapid reconnection. A
similar scenario was also proposed in the flare events by Wang \&
Shi (1993). They suggested that there is a slow reconnection between
two topologically separated loops in the lower atmosphere and the
slow reconnection triggers the fast reconnection in the corona which
is responsible for large solar activities.

\acknowledgments { This work is supported by the National Natural
Science Foundations of China (11203037, 11221063, 11303049, and
11373004), the Outstanding Young Scientist Project 11025315, the CAS
Project KJCX2-EW-T07, the National Basic Research Program of China
under grant 2011CB811403, and the Strategic Priority Research
Program$-$The Emergence of Cosmological Structures of the Chinese
Academy of Sciences (No. XDB09000000). The data are used by courtesy
of NVST, AIA, and HMI science teams. }

{}

\clearpage

\end{document}